\documentclass[aip,jcp,reprint,floatfix]{revtex4-1}
\usepackage{verbatim}%
\usepackage{amsmath,amsfonts} %
\usepackage[acronym]{glossaries} %
\usepackage{graphicx} %
\usepackage{dcolumn} %
\usepackage{bm} %
\usepackage{ifpdf} %
\usepackage{color} %
\usepackage{nicefrac} %
\usepackage{natbib} %
\usepackage[byname]{smartref} %
\usepackage[breaklinks, %
colorlinks, %
linkcolor=blue, %
citecolor=blue, %
urlcolor=blue]{hyperref} %
\usepackage[table]{xcolor} %
\usepackage{braket} %

\newcommand{\eq}[1]{Eq.~(\ref{#1})} %
\newcommand{\bea}{\begin{eqnarray}}
\newcommand{\eea}{\end{eqnarray}}

\ifpdf %
 \fi

\begin{document}
                            
\newacronym{CI}{CI}{conical intersection} %
\newacronym{GP}{GP}{geometric phase} %
\newacronym{BO}{BO}{Born-Oppenheimer} %
\newacronym{LVC}{LVC}{linear vibronic coupling} %
\newacronym{DOF}{DOF}{degrees of freedom} %
\newacronym{PES}{PES}{potential energy surface} %
\newacronym{DBOC}{DBOC}{diagonal Born--Oppenheimer correction} %
\newacronym{BMA}{BMA}{bis(methylene) adamantyl} %
\newacronym{FC}{FC}{Franck-Condon} %
\newacronym{CWE}{CWE}{cylindrical wave expansion}%
\newacronym{PB}{PB}{Particle in a box}%
\newacronym{BOA}{BOA}{Born-Oppenheimer approximation} %
\newacronym{BSC}{BSCs}{bound states in the continuum}

\title{Topological origins of bound states in the continuum for systems with conical intersections}

\author{Sarah Henshaw} %
\author{Artur F. Izmaylov} %
\affiliation{Department of Physical and Environmental Sciences,
  University of Toronto Scarborough, Toronto, Ontario, M1C 1A4,
  Canada} %
\affiliation{Chemical Physics Theory Group, Department of Chemistry,
  University of Toronto, Toronto, Ontario M5S 3H6, Canada} %

\date{\today}

\begin{abstract}
\Gls{BSC} were reported in a linear vibronic coupling model with a \gls{CI} [Cederbaum {\it et al.} Phys. Rev. Lett. 
\textbf{90}, 013001 (2003)]. It was also found that these states are destroyed within the \gls{BOA}.
We investigate whether a nontrivial topological or \gls{GP} associated with the \gls{CI}
is responsible for \gls{BSC}. To address this question we explore modifications of 
the original two-dimensional two-state linear vibronic coupling model supporting \gls{BSC}.
These modifications either add \gls{GP} effects after the \gls{BOA} 
or remove the \gls{GP} within a two-state problem. Using the stabilization graph technique we shown that 
the \gls{GP} is crucial for emergence of \gls{BSC}. 

\end{abstract}

\pacs{}

\maketitle

\glsresetall



{\it Introduction:} In most quantum mechanical problems, the bound states emerged in the continuum 
of unbound states become resonances with a finite lifetime when their interaction with the continuum 
is accounted. However, from the dawn of quantum theory, von Neumann and Wigner\cite{Neuman:1929/physz} 
discovered potentials that support spatially bounded, discrete states with energies within the continuum. 
Later, such \gls{BSC} were found not only in the time-independent Schr\"odinger equation but also in 
the wave equation,\cite{Capasso:1992,Soljacic/nature/2013,kante:nature/2017,Moiseyev:2009es,Bogdanov:2017/acsphot} 
where they gave rise to nanophotonic applications in lasing, sensing, and filtering.\cite{wei:2016rev}     

  
In 2003, \Gls{BSC} were also found in a nonadiabatic model relevant to molecular dissociation.\cite{Cederbaum:2003/prl/013001}  Cederbaum {\it et al.} considered a two-dimensional two-state linear vibronic 
 coupling Hamiltonian\cite{Cederbaum:2003/prl/013001}
\begin{equation}
  \label{eq:Hdiab}
  H_{\rm dia} = { T_N}  {\mathbf 1}_2 + 
  \begin{pmatrix} 
    V_{b} & V_{bc} \\
    V_{bc} & V_{c}
  \end{pmatrix},
\end{equation}
where $ T_N$ = $-\frac{1}{2} (\omega_x \partial^2 _x + \omega_y \partial^2_y )$ 
is the nuclear kinetic energy operator (atomic units are used throughout),  
$V_{b}$ and $V_c$ are the bound and unbound potentials 
\begin{align}
  \label{eq:diab-me-11}
  V_{b} = {} & \frac{\omega_x }{2}x^2
  + \frac{\omega_y }{2}y^2,  \\
  \label{eq:diab-me-c}
  V_{c} = {} & \epsilon e^{-\beta \left(x+\delta \right)} +
  \frac{\omega_y}{2}y^2 
\end{align}
coupled by a linear potential $V_{bc}=\lambda y$. The coordinates
$x$ and $y$ can be thought as mass and frequency weighted, 
the parameters were set to
 $\omega_x$= 0.015, $\omega_y$=0.009, $\lambda = 0.01$, 
 $\epsilon$ = 0.04, $\beta$ = 0.5, and $\delta$ = 0.5. 
 Even though all vibrational states of the $V_b$ are coupled with continuum states 
 of $V_c$, it was discovered that the lowest state of the $V_b$ gives rise to a BSC.
 A simple way to understand this result is to inspect what continuum states 
 can be coupled with the ground state of $V_b$. Due to the linear dependence of $V_{bc}$ and separable 
 $x$ and $y$ components of $V_c$, all such states can be denoted as $(k,1)$, where $k$ is the quantum number 
 along the $x$-direction and $1$ is the vibrational quantum along the $y$-direction. Comparing the energy 
 of the $V_b$ ground state, $(\omega_x+\omega_y)/2 = 0.012$, with the lowest energy of the $(k,1)$ manifold, 
 $3~\omega_y/2 = 0.0135$, a gap that makes the bound ground state off-resonance from states of 
 the coupled $(k,1)$ continuum becomes evident. This gap lifts the edge of the coupled continuum above the energy
 of the bound state and thus effectively breaks the state emergence in the continuum.  
 %
 \begin{figure}
  \includegraphics[width=\linewidth]{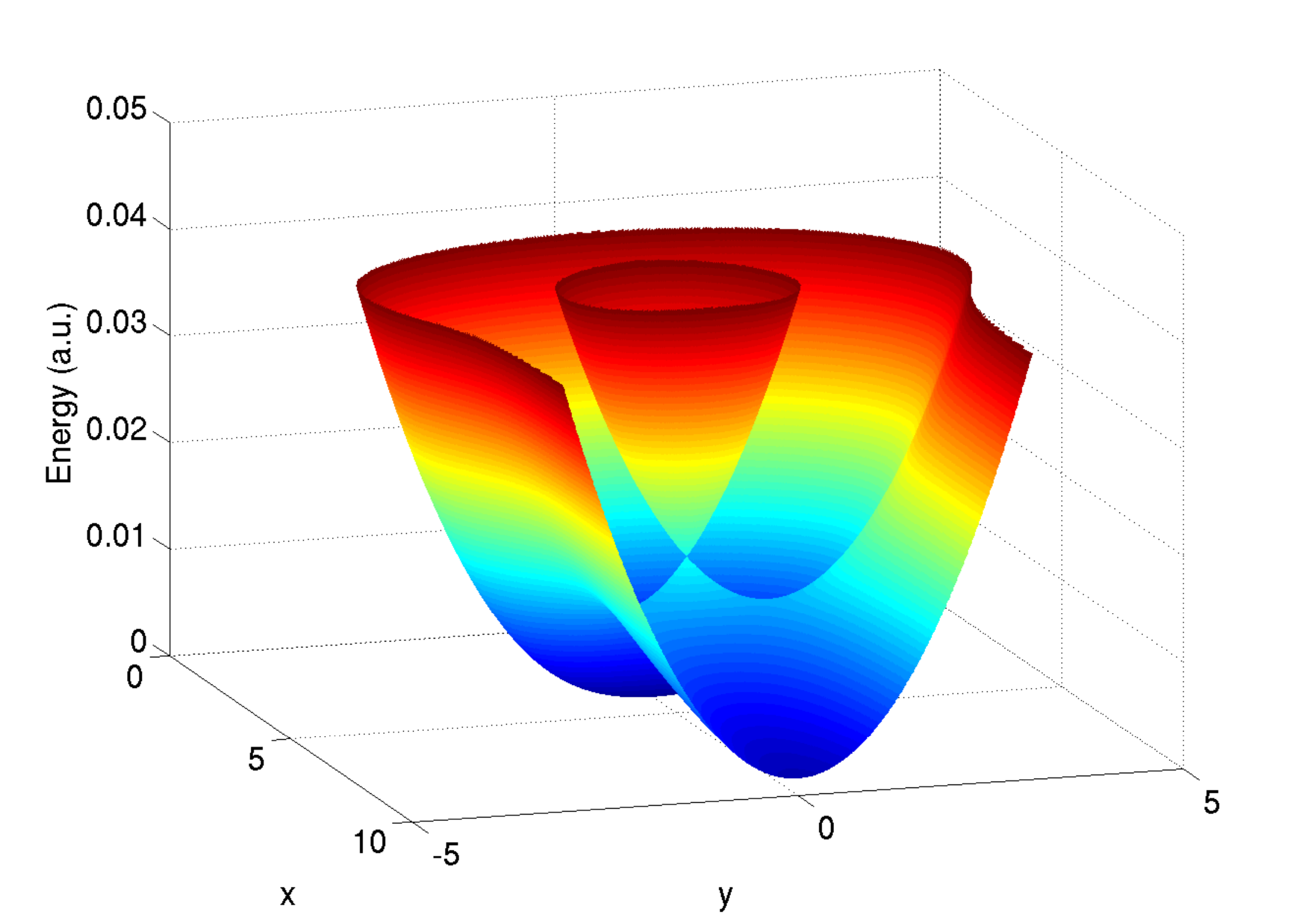}
  \caption{Model potentials in the adiabatic representation.}
  \label{fig:adiab3D}
\end{figure}
 
Transforming the problem to the adiabatic representation gives rise to a CI (Fig.~\ref{fig:adiab3D}).
Considering that CI's energy ($E_{\rm CI} \approx 0.015$) 
is higher than the energy of the ground state of the $V_b$ potential ($0.012$), one can conclude that 
the \gls{BOA} may be quite adequate at least for this state. It turns out that switching to the \gls{BOA}
destroys the bound state and gives rise to a resonance state.  
This shows that purely energetic consideration is not accurate in this case and some
symmetry is broken when the \gls{BOA} is introduced.  
   
Generally, \glspl{CI} not only  promote
transitions between different electronic states but also introduce Berry or \gls{GP}.\cite{LonguetHigg:1958/rspa/1,Berry:1984/rspa/45,Izmaylov:ACR/2017} 
The manifestation of \gls{GP} is in changing a sign of the electronic wavefunction upon 
a continuous evolution along a closed path encircling a \gls{CI}.
To ensure that the total wavefunction is single-valued, the nuclear wavefunction must also change the sign. 
This sign change can be introduced via a phase factor that is position-dependent and constitutes exponential 
function of the \gls{GP}. 
\gls{GP} is a feature of the adiabatic representation and does not appear in the diabatic representation. 
In several instances \gls{GP} played a crucial role in obtaining qualitatively correct results when 
 problems were considered in the adiabatic representation\cite{Ryabinkin:2013/prl/220406,Kendrick:1997/prl/2431,Ham:1987/prl/725,Guo:JACS/2016,YarkonyGuo/2017/JCTC,Xie:2017hk,Xie:2017ja}.
Therefore, it is quite natural to inquire whether the \gls{GP} 
is the reason for appearance of \gls{BSC} in this system. 
 
{\it Methods:} To investigate the role of the GP we will consider time-independent Schr\"odinger equation for four 
Hamiltonians. The original diabatic Hamiltonian $H_{\rm dia}$ [\eq{eq:Hdiab}] is used as a reference that includes
all contributions: nonadiabatic transitions and GP effects. In contrast, the \gls{BO} Hamiltonian $H_{\rm BO}$ does not 
have any of these effects. Formally, to obtain $H_{\rm BO}$ one needs to diagonalize the potential matrix in 
$H_{\rm dia}$ using the unitary matrix
\begin{equation}
  U = 
  \label{eq:Umat}
  \begin{pmatrix}
    \cos\theta & \sin\theta \\
    -\sin\theta & \cos\theta
  \end{pmatrix}
\end{equation}
where angle $\theta$ is  
\begin{equation}
  \label{eq:theta}
  \theta(x,y) = \frac{1}{2}\arctan \dfrac{2\,V_{bc}}{V_{c} - V_{b}} .
\end{equation}
Removing the high energy electronic state and nonadiabatic couplings gives
\begin{equation}
H_{\rm BO} = T_N + W_{-},
\end{equation}
where 
\begin{equation}
W_{-} = \frac{1}{2} \left(V_{b}+V_{c} \right) - \frac{1}{2} \sqrt{\left(V_{b}-V_{c}\right)^2+4 V_{bc}^2}.
\end{equation}

To include \gls{GP} in the \gls{BO} representation we use the Mead and Truhlar approach\cite{Mead:1979/jcp/2284},
where the double-valued projector $e^{-i \theta}$ is applied to $H_{\rm BO}$
to avoid working with double-valued nuclear wavefunctions.  
This results in our third Hamiltonian 
\bea \notag
H^{\rm GP}_{\rm BO} &=& e^{ i \theta} H_{\rm BO} e^{-i \theta} \\
&=&T_N + \tau_{\rm GP} + W_{-},
\eea
where
\begin{equation}
\tau_{\rm GP} = \frac{\left(\nabla \theta \right)^2}{2} + i \nabla \theta \nabla + \frac{i}{2} \nabla^2 \theta,
\end{equation}
and $\nabla = (\sqrt{\omega_x}\partial_x, \sqrt{\omega_y}\partial_y)$.
$H^{\rm GP}_{\rm BO}$ includes only the GP effects but excludes all nonadiabatic transitions. 

Finally, to obtain a picture 
where nonadiabatic transitions 
are preserved but GP effects are removed, we use the diabatic Hamiltonian identical to $H_{\rm dia}$ but with 
modified $V_{bc} = c|y|$. Introducing the absolute value function in the coupling was shown to remove 
the GP when the diabatic-to-adiabatic transformation is done.\cite{IzmaylovJiaru:2016/acs}
We will refer to this Hamiltonian as $H_{\rm dia}^{\rm noGP}$.
An alternative to $H_{\rm dia}^{\rm noGP}$ can be a Hamiltonian obtained by 
transforming $H_{\rm dia}$ to the adiabatic representation and ignoring the double-valued 
boundary conditions introduced by this transformation. The main advantage of $H_{\rm dia}^{\rm noGP}$
is numerical robustness of the diabatic representation. 

To determine if an eigenstate is a bound or resonant state,
 its lifetime is calculated using the stabilization method.\cite{PhysRevLett.70.1932} 
Stabilization graphs are used to obtain complex energies 
in the form $E = E_R-i \Gamma /2$, where $E_R$ is the real part and $\Gamma$
 is inversely proportional to the lifetime of the state.\cite{Book/Nimrod:2011}
A finite lifetime of the state is an indication of its resonance character. 
\begin{figure}
  \includegraphics[width=\linewidth]{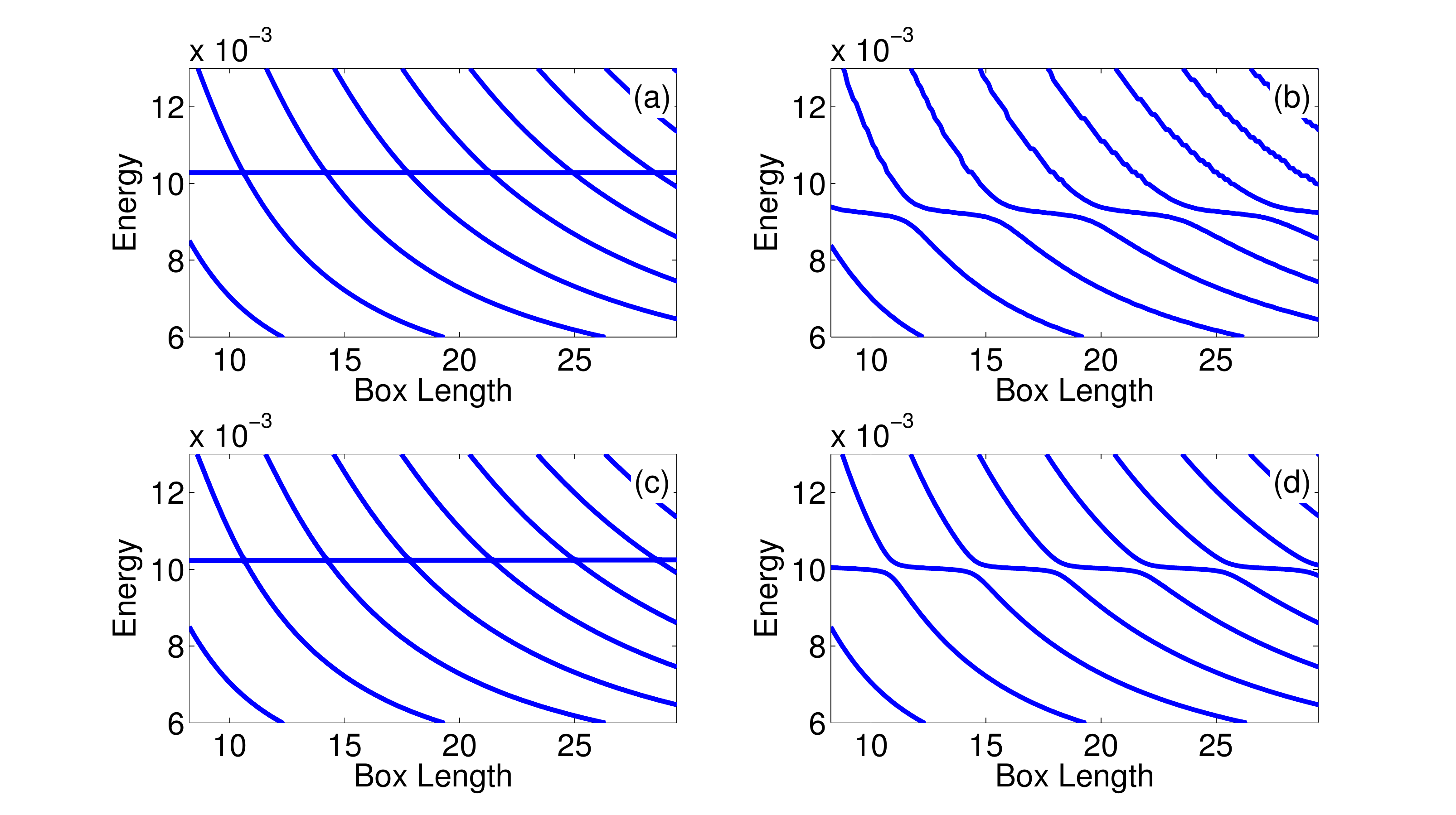}
  \caption{Stabilization graphs for the lowest localized state of various Hamiltonians: a) $H_{\rm dia}$, 
  b) $H_{\rm BO}$, c) $H_{\rm BO}^{\rm GP}$, and d) $H_{\rm dia}^{\rm noGP}$.}
  \label{fig:stab}
\end{figure}

All Hamiltonians are transformed to matrices using static basis functions.
The same product basis of functions in $x$ and $y$ directions was used 
for the two electronic states of the diabatic Hamiltonians as well as for the single electronic state adiabatic 
Hamiltonians. \gls{PB} eigenfunctions were used in the $x$-direction. These functions introduce the length of the box 
as a natural stabilization parameter for the stabilization method.\cite{PhysRevLett.70.1932}
The box interval $[-4,L]$ chosen to cover both bound and continuum parts of the potential 
(Eqs.~(\ref{eq:diab-me-11}) and (\ref{eq:diab-me-c})).  
Harmonic oscillator eigenfunctions were used in the $y$-direction. Energy of low lying states for all Hamiltonians
converged within $10^{-6}$ a.u. with 150 \gls{PB} and 20 harmonic oscillator eigenfunctions. 

\begin{figure}
  \includegraphics[width=\linewidth]{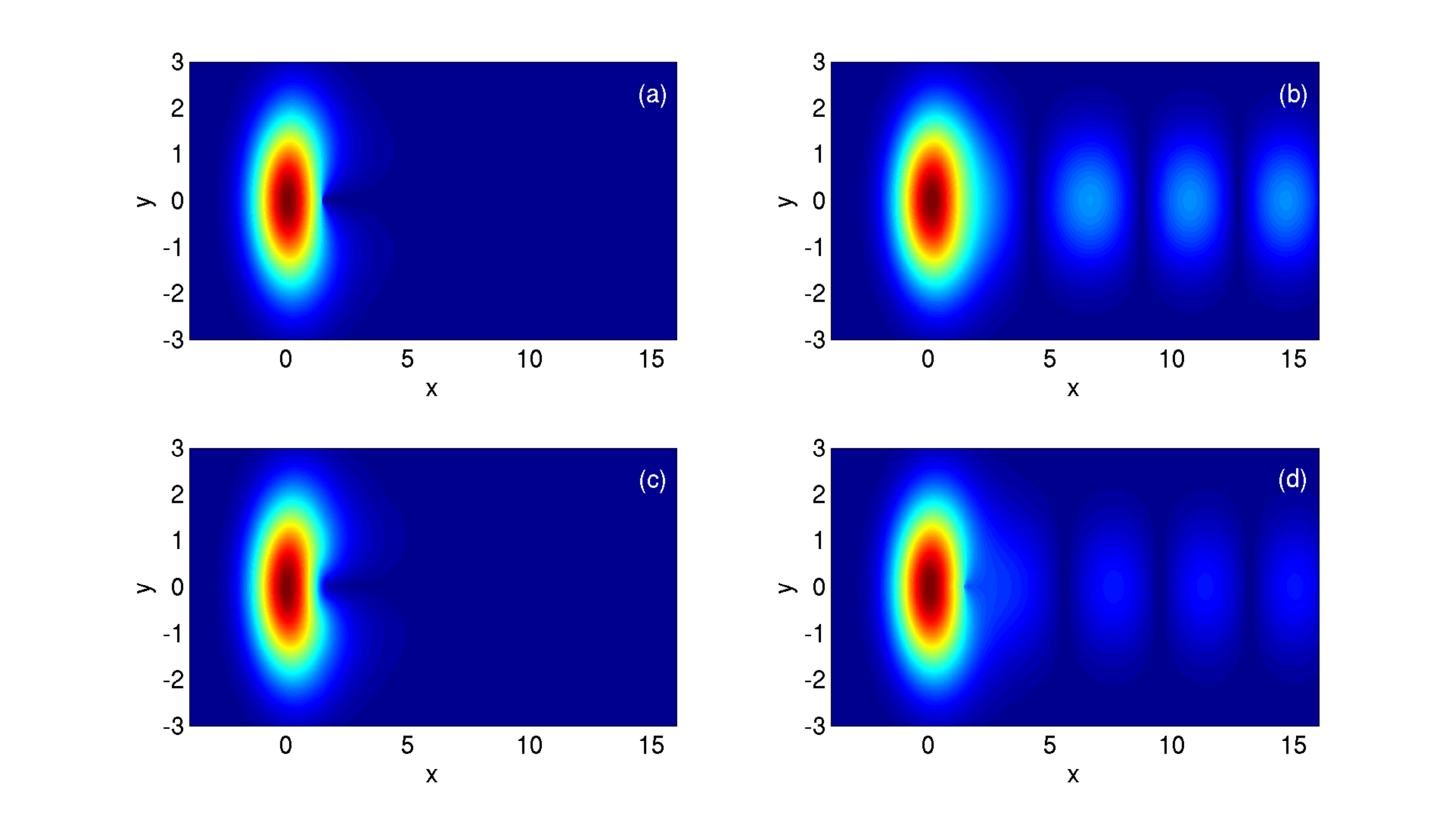}
  \caption{Contour of the square root of the ground electronic state nuclear probability density 
  of the lowest localized state for different Hamiltonians 
  with box length $20.4$ a.u.:  
   a) $H_{\rm dia}$, b) $H_{\rm BO}$, c) $H_{\rm BO}^{\rm GP}$, and d) $H_{\rm dia}^{\rm noGP}$.}
  \label{fig:contour}
\end{figure}

{\it Results:}  Table \ref{tab:en} and Figs.~\ref{fig:stab} and \ref{fig:contour} 
summarize results of stabilization calculations
for the lowest localized states of the four Hamiltonians. The Hamiltonians accounting for GP effects
have only real components of energy up to a numerical error, while removing GP effects leads to 
a resonance character of the lowest localized state. Thus, the BSC can be clearly related to the presence of the GP. 
An intuitive picture of this relation is that in the absence of the GP the nuclear wave-packet can tunnel through 
a barrier separating the bound part of the $W_{-}$ potential from its unbound counterpart, whereas 
addition of the GP creates destructive interference between the parts of the wave-packet that tunnel 
through the barrier on different sides of the CI. 
 \begin{table}[!h]
  \caption{The energies of the lowest localized state of different Hamiltonians (in $10^{-3}$ a.u.).} 
  \label{tab:en}
  \centering
  \begin{ruledtabular}
    \begin{tabular}{@{}cccc@{}}
    $H_{\rm dia}$ & $H_{\rm BO}$ & $H_{\rm BO}^{\rm GP}$ & $H_{\rm dia}^{\rm noGP}$  \\ \hline
      $10.29$ & $9.23-i 0.10$ & $10.23$ & $10.02-i0.04$ \\ [1ex] 
    \end{tabular}
  \end{ruledtabular}
\end{table}
The absolute overlap of the $H_{\rm dia}$ and  $H_{\rm BO}^{\rm GP}$ wavefunctions, 
$|\langle\Psi_{\rm dia}|\Psi_{\rm BO}^{\rm GP}\rangle|$ is 0.93, which is relatively large considering  
that the total probability to find the system described by $|\Psi_{\rm dia}\rangle$ 
in the ground electronic electronic state is 0.97 (see also Fig.~\ref{fig:contour} (a) and (c)).  


In conclusion, we have unambiguously shown that the nontrivial \gls{GP} is the reason for  
bound states in the continuum in the considered model problem. This result 
is yet another illustration that inclusion of the GP can qualitatively change results in 
nonadiabatic problems and cannot be ignored {\it a priori}. The current result is a direct 
analog of GP induced localization obtained in the double-well potential problem\cite{Ryabinkin:2013/prl/220406,JoubertDoriol:2017if}.
As in the double-well case, one can expect that there is a certain range of parameters of the linear 
vibronic model that supports the BSC. This is in accord with previous works on GP effects in tunneling 
that did not find BSCs for a very similar model with different values of corresponding 
parameters\cite{Xie:2017hk,Xie:2017ja}.     
Yet, one can hope that it is possible to engineer a molecular system where GP will not only slow down the tunneling 
process but will completely freeze it by giving rise to BSCs.


{\it Acknowledgements:} The authors thank Ilya Ryabinkin and Lo{\"i}c Joubert-Doriol for helpful discussions.
This work was supported by a Sloan Research Fellowship, Natural Sciences and 
Engineering Research Council of Canada (NSERC), and an Ontario Graduate 
Scholarship (OGS).
%

\end{document}